\documentclass[12pt]{iopart}

\usepackage{iopams}  
\usepackage[upright]{fourier}
\usepackage{graphicx}
\usepackage{dcolumn}

\begin{document}

\title{Pulsed voltage cold atmospheric plasma jet and gold nanoparticles enhance cytotoxic anticancer effect}

\author{I Schweigert$^{1,\star}$, M Biryukov$^{1,2}$, A Polyakova$^{1,2}$, N Krychkova$^{1,2}$, E Gorbunova$^{1,2}$, A Epanchintseva$^2$, I  Pyshnaya$^{2}$, Dm~Zakrevsky$^{1,3}$, E Milakhina$^{1,3}$, O Koval$^{1,2}$}

\address{$^1$Khristianovich Institute of Theoretical and Applied Mechanics, Novosibirsk, Russia}
\address{$^2$Institute of Chemical Biology and Fundamental Medicine, Novosibirsk, Russia}
\address{$^3$Novosibirsk State Technical University, Novosibirsk, Russia}
\ead{$^\star$ischweig@yahoo.com}

\vspace{10pt}
\begin{indented}
\item[]August 2023
\end{indented}

%\maketitle

\begin{abstract}
Efficient and biologically safe mode of cold atmospheric plasma jet (CAPJ) is crucial for the development of CAPJ-based anticancer therapy. In the experiment and numerical simulations, by changing the pulse duration of a positive-pulsed voltage, we found the optimal CAPJ mode with regular streamer propagation. CAPJ regimes with a maximum discharge current at a temperature $\it T$$<$42$^\circ$C substantially suppressed the viability of cancer cells. To enhance cell killing, gold nanoparticles (NPs) were added to the cells before and after the CAPJ exposure. Combination of CAPJ, generated with positive-pulsed voltage, and gold nanoparticles decreased viability of NCI-H23 epithelial-like lung adenocarcinoma, A549 lung adenocarcinoma, BrCCh4e-134 breast adenocarcinoma and uMel1 uveal melanoma cells. Polyethylene glycol-modified nanoparticles with attached fluorescent label were used to visualize the uptake of NPs. We demonstrated that NPs efficiently entered the cells when were added to the cells just before CAPJ exposure or up to two hours afterwards. The efficiency of NPs penetration into cells positively correlated with the induced cytotoxic effect: it was maximal when NPs was added to cells right before or immediately after CAPJ exposure. 
Summarizing, the treatment with optimal CAPJ modes in combination with modified NPs, bearing the cancer-addressed molecules and therapeutics may be next strategy of strengthening the CAPJ-based antitumor approaches.

%A combination of cold atmospheric  plasma jet (CAPJ), generated with positive-pulsed voltage, and gold nanoparticles (NPs) was used to treat cell lines of  NCI-H23 epithelial-like lung adenocarcinoma, A549 lung adenocarcinoma, BrCCh4e-134 breast  adenocarcinoma, and uMel1 uveal melanoma cells. In the experiment and numerical simulations, by changing the pulse duration of a  positive-pulsed voltage, the optimal CAPJ mode with regular streamer propagation is found. Optimal mode is determined by measuring the discharge current and temperature near the dielectric plate and shaved mouse skin. CAPJ regimes with a maximum discharge current at a temperature T$<$42$^\circ$C, which is safe for biological tissues, are used to suppress the viability of cancer cells.  To enhance the efficiency of CAPJ exposure, gold nanoparticles  are added to the medium with cells at different time intervals before or after CAPJ irradiation. We use polyethylene glycol-modified nanoparticles with attached fluorescent labels to visualize the uptake of NPs.  The viability of cancer cells  after co-treatment with optimal regimes of CAPJ and NPs is discussed, as well as the correlation between NPs uptake and cell death.
\end{abstract}
%\keywords{Cold atmospheric plasma jet, positive-pulsed voltage, gold nanoparticles,  anticancer therapy, lung adenocarcinoma}
%\submitto{\jpg}
\maketitle
%\end{verbatim}
\normalsize

\section{Introduction}

Low-temperature plasma anticancer therapy is an actively developing field (see, for example, reviews \cite{Zivani2023,Limanowski} and references therein).  
Despite the different designs and performance characteristics of DBD and plasma jet sources, the treatment of cancerous lesions 
with cold atmospheric plasma has shown promising results in research and clinical trial \cite{Lya2023}. 
CAPJ treatment is cytotoxic to tumors at high doses (high voltage and long exposure time) and stimulating for wound healing at low doses (a low energy input and a short exposure time) 
\cite{Woedtke}.
Therefore, plasma sources operating in "high-dose" modes with low heating of biological tissue are more suitable for anticancer therapy. 

Most plasma sources that generate CAPJ for  medical applications work with sinusoidal voltages. 
Required characteristics such as a large energy input, low heat, and strong electric field in the streamer head approaching the target surface are not easily achieved with  sinusoidal CAPJ excitation.
Streamers in the high-voltage CAPJ not only  generate more radicals in the surrounding air but also deliver a large electric field to the target, which temporarily changes the permeability of the cancer cell membrane \cite{Yan2018}. 
The values  of temperature and voltage  can be decoupled when the CAPJ is ignited by a positive-pulsed voltage with a trapezoidal pulse shape \cite{SCH2022,SCH2023}.
The thermal effect at high voltage can be reduced by changing the duration of the voltage pulse.
Due to this advantage, CAPJ excited by 
positive-pulsed voltage is more preferable for medical applications.
To enhance the efficacy of plasma anticancer therapy this novel approach is often used with traditional chemo-, immuno- and nanotherapies. 
Synergism in enhancing the cytotoxicity of cold gas discharge plasma and gold nanoparticles was first shown by Kim ${\it et}$ ${\it al.}$  \cite{Kim2009}. 
Later, cold gas discharge plasma combined with gold nanoparticles was successfully used in numerous  cancer studies in vitro \cite{Kim2011,Cheng2014,Choi2015,Irani2015,Choi2017,Kim2017,Jawaid2020} and in vivo \cite{Kaushik2016}. 
The gold nanoparicles pure and covered with polyethylene glycol (PEG) are biocompatible and can be employed for drug/antibody delivery.
Surface receptors overexpressed in cancer cells  are often used as target molecule for the conjugation with a drugs or drug-bearing nanoparticles \cite{Al-Harbi2020}.
 The gas discharge plasma enhances nanoparticle delivery to cells and increases the concentration of reactive oxygen and nitrogen species in target tissues \cite{Schmidt,Lademann,OLademann,OLademann1}. The electric field generated by gas discharge plasma is one of important factors for efficient transport and penetration of biological materials and nanoparticles into cells \cite{Cheng2014,Busco,Kaneko,Shaw}.

The effect of combined exposure to gas-discharge plasma and gold nanoparticles
has often been studied on glioblastoma cells.
This combined treatment suppressed cell proliferation rate, increased NPs uptake and decreased cancer cell migration
\cite{Cheng2014,Kaushik2016,He2018,Cheng2015,He2020}. 
Similar results were obtained for melanoma when cold gas discharge plasma and nanoparticles were used together 
\cite{Kim2009,Choi2015,Choi2017,Jawaid,Bekeschus2021}.

In this work, we use high-voltage helium cold atmospheric plasma jet and gold nanoparticles to kill  cancer cells. Our new step in the field of plasma-nanoparticle co-treatment is the application of high-voltage  modes of CAPJ  ignited by positive-pulsed (PP) or sinusoidal voltages. 
The CAPJ operation regimes are characterized using the measured and calculated discharge current on the target and the temperature in a zone plasma-target contact. The optimal CAPJ regime is the mode with the maximum discharge current near the target, with regular streamer propagation and minimal heating of the target. To obtain this optimal regime we vary the pulse duration of the positive-pulsed voltage.

The paper is organized as follows. 
The methods and materials are given in section 2. The CAPJ regimes are described in section 3. An analysis of the sensitivity of cancer cells to the combined effects of CAPJ and NPs is presented in section 4. The results of internalization of NPs with CAPJ are discussed in section 5.  The conclusions are in section 6.

\section{Methods and materials} 

\subsection{Plasma source}
The plasma jet source in Fig.\ref{setup} is a discharge cell in the form of a coaxial dielectric channel with a length of 10 cm and an inner diameter of 1 cm. In the center of the channel there is a copper electrode in the form of a rod with a length of 5 cm and a diameter of 0.2 cm, to which a voltage $\it U$ is applied. A dielectric capillary with an inner diameter of 0.23~cm and a length of 0.5~cm was placed in the nozzle. 
In the experiments the cross-sectional area of the free space for gas flow is 0.75~cm$^2$.
The discharge zone  formed between the powered electrode and an annular grounded electrode located on a dielectric tube near the nozzle. A detailed description of the plasma source can be found in Refs.{\cite{SCH2019,SCH2020L}. 

The helium plasma jet was initiated with a generator of sinusoidal voltage with the amplitude $\it U$=3.3 -- 3.5~kV at a fixed frequency ${\it f}_U$ = 50 kHz and with a generator of unipolar positive pulses with adjustable repetition rate f = 1 -- 40 kHz and amplitude of 3.8~kV and 4.2~kV. The pulse duration $\it \tau$ varied from  5~$\mu$s to 16~$\mu$s.
The voltage amplitude $\it U$ was limited to ensure safe conditions of CAPJ exposure to living objects. 
An ohmic high impedance divider was used for voltage measurements. Current measurements were carried out by a sensor located at a distance z=2.5 cm from the nozzle perpendicular to the axis of plasma jet propagation and representing a collector (flat metal electrode). 
The collector was grounded through a shunt (low inductive resistance).  This allowed measuring the frequency and amplitude of the current pulse on the collector.
In all our experiments, a grounded metal collector was used as an additional electrode to enhance an electric field in the streamer head. This led to the intensification of active  radical generation in the contact zone of the CAPJ with the target compared to the case with a target under a floating potential. Previously we studied the effect of presence of grounded electrode beneath the target (see Refs.\cite{SCH2019,SCH2022F}). The distance from the nozzle to the target of 2.5 cm was chosen so that the plasma jet touched the target, providing maximum OH generation at a given gas flow rate and minimum heating of the target.
Three types of targets were used in the experiments: a 0.1 cm thick Al$_2$O$_3$ plate as a test target, cells in media and mice on the dielectric substrate. In all cases the grounded electrode was beneath the target.  The helium flow rate was 9 L/min. Our earlier experiments \cite{SCH2019} demonstrated that for the geometry of the plasma jet source used in this study, a  laminar gas flow regime was observed for the flow rate $\it v$ up to 12 L/min. With further increase of $\it v$ ,  the flow becomes unstable. We also showed that the maximum luminescence intensity of the OH radical ($\lambda $=309 nm) was observed at $\it v$ =6 -- 9 L/min.

\begin{figure}
\centering
\includegraphics[width=5.3 cm]{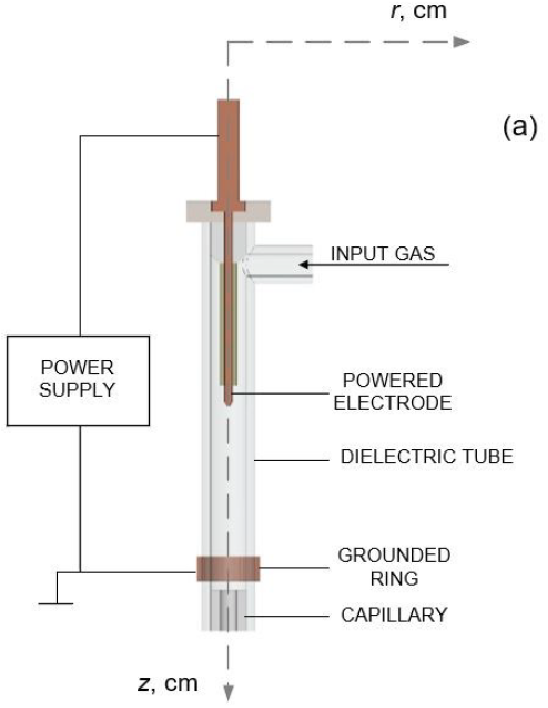}
\includegraphics[width=4.5 cm]{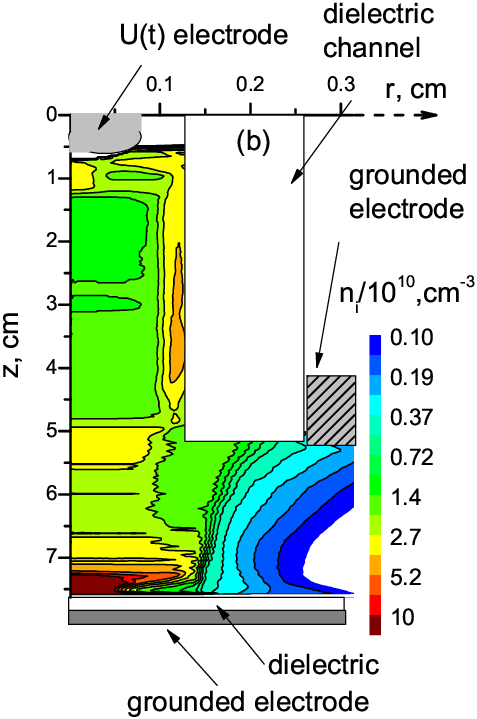}
\caption{
(a) Scheme of experimental device for generation of cold atmospheric plasma
jet  and (b) calculation domain with the ion density distribution, when a streamer touching the target, for positive-pulsed voltage with the amplitude of 3.8 kV and frequency of 30 kHz.
 }
\label{setup}
\end{figure}
\subsection{Simulation model} 
  
  The numerical simulation of the discharge ignition and streamers dynamics for the  conditions of experiment was carried out with the fluid model approach with 2DPlasmaNovH code. Gas discharge plasma model and simulation details were presented in Ref. \cite{SCH2019_1}. In this study, for an analysis of streamer propagation patterns (a combination of short and long streamers in plasma jet), the dynamics of streamers was simulated for several dozen voltage pulses. Therefore, we used a simplified model that considered only helium gas jet without nitrogen or oxygen admixture. In most of our experiments, the helium gas velocity is   30~m/s ($\it v$ = 9~L/min), that justifies our assumption. The formation of molecular ions and Penning processes are not taken into account in the model, since we assume that the high-rate helium gas flow refreshes gas between the nozzle and the target. The good agreement between the calculated and measured discharge currents at the surface, and the streamer propagation characteristics confirmed the correctness of our assumptions. The comparison of the experimental and computed data, as well as the discussion on the results are given in the section 3. For the photoionization calculation we used the model developed in \cite{SCH2019_1}. In our experiments \cite{SCH2019_1}, the streamer images were glowing spots of approximately 0.2 -- 0.25 cm in radius around the streamer head. From these observations, it followed that the characteristic time of the emission of photons by excited atoms was smaller than the characteristic time of the streamer propagation, and that the characteristic length of photoionization was less than 0.25 cm. The model assumptions based on the experimental observations are (a) photons are emitted instantly after the atom excitation and  (b) the generation of electron/ion pairs by these photons takes place within an area around the streamer head.
  In the calculation domain (Fig.\ref{setup} (b)), the distance between the powered electrode and the ring grounded electrode is 4.5 cm, and the distance between the nozzle and the dielectric plate is 25 mm. 
The radius of the powered electrode located in the dielectric channel is 0.1 cm. 
The simulation domain is cylindrical with radius $\it R$ = 7 cm and the height $\it Z$ = 7.8~cm. 

\subsection{Gold  nanostructures}

Gold nanoparticles of 7 and 13 nm in diameter were synthesized by methods from Refs.\cite{Jana2001,Murphy2004} and covered with polyethylene glycol (PEG) according to Ref.\cite{Bartczak}.
 The polyethylene glycol derivative contained thio (SH) group which allows covalent attachment of PEG to NPs  and carboxyl groups (COOH) can be used for covalent attachment of a fluorescent label or guide molecule to NPs. 
The fluorescein amidite (abbreviated as FAM) was attached as a label to visualize the penetration of the entire nanostructure into cells.
The physicochemical characteristics of  NPs  and NP-PEG shown in Fig. \ref{k3n} were obtained with gel electrophoresis, transmission electron microscopy with a JEM-1400 Jeol microscope, spectrophotometric method with a UV-2100 spectrophotometer (Shimadzu, Japan), and by dynamic light scattering (DLS) with a Zetasizer Nano ZS. 
\begin{figure*}
%\begin{figure}[ht!]
	\centering
\includegraphics[width=6 cm]{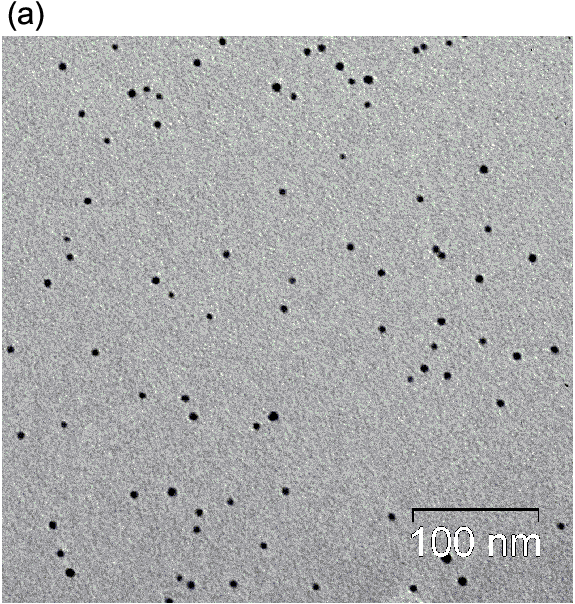}
\includegraphics[width=6 cm]{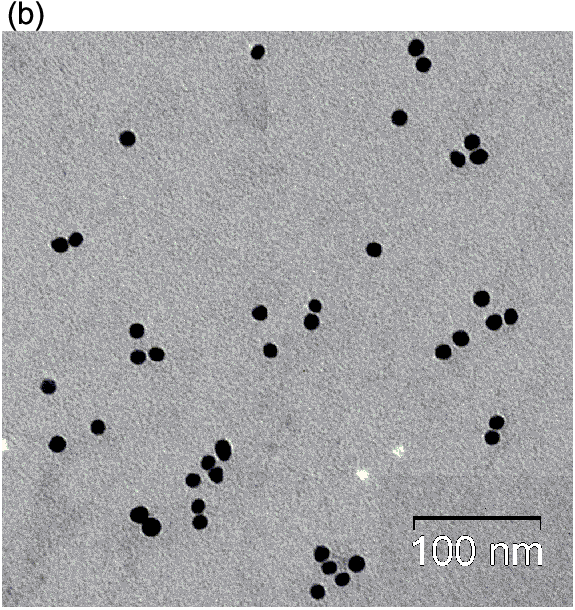}
\includegraphics[width=6 cm]{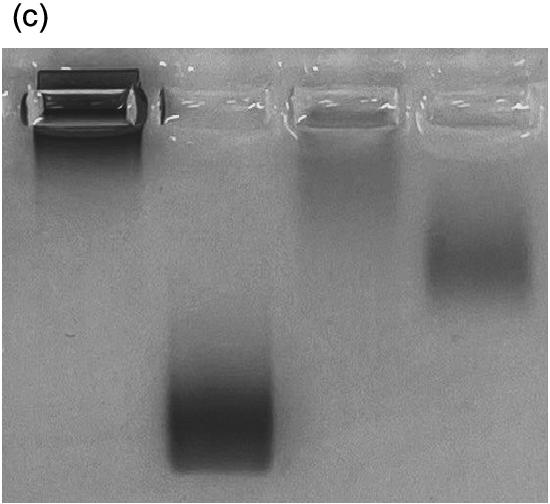}
\includegraphics[width=7 cm]{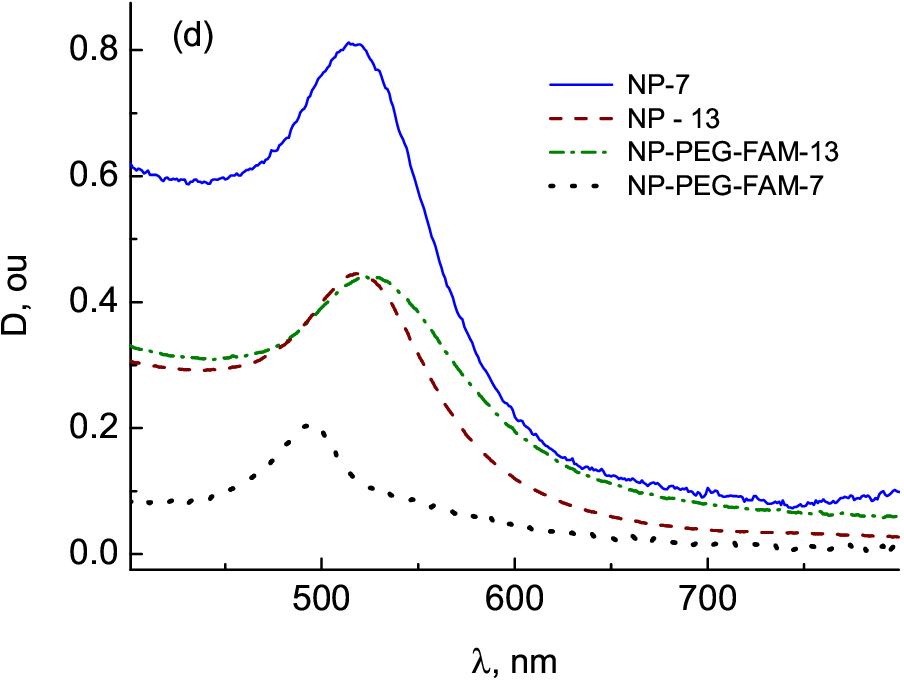}
\includegraphics[width=7 cm]{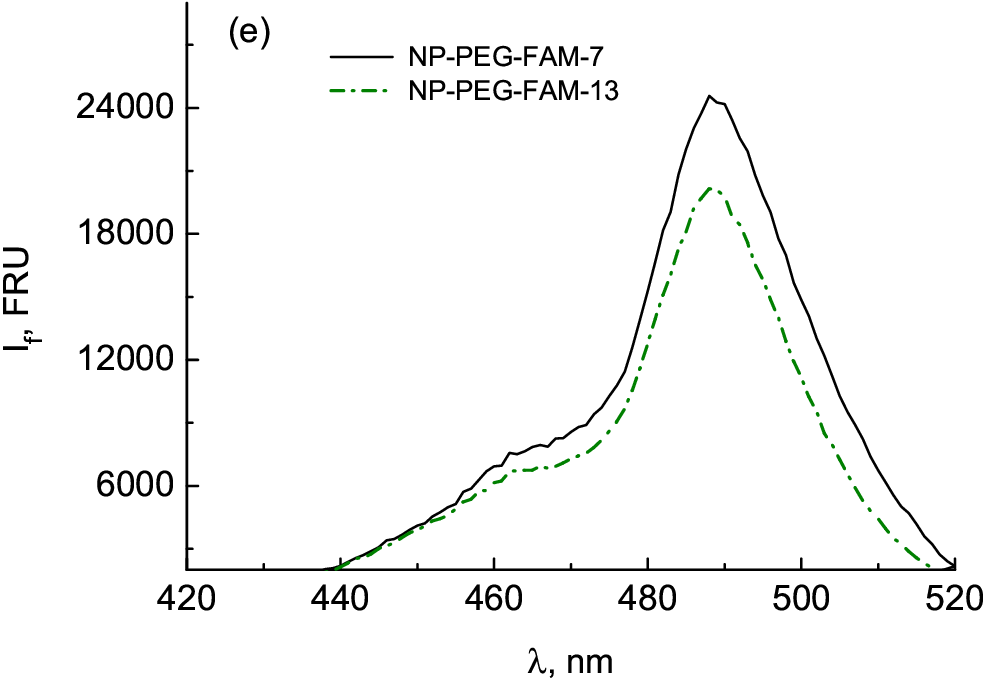}
\includegraphics[width=7 cm]{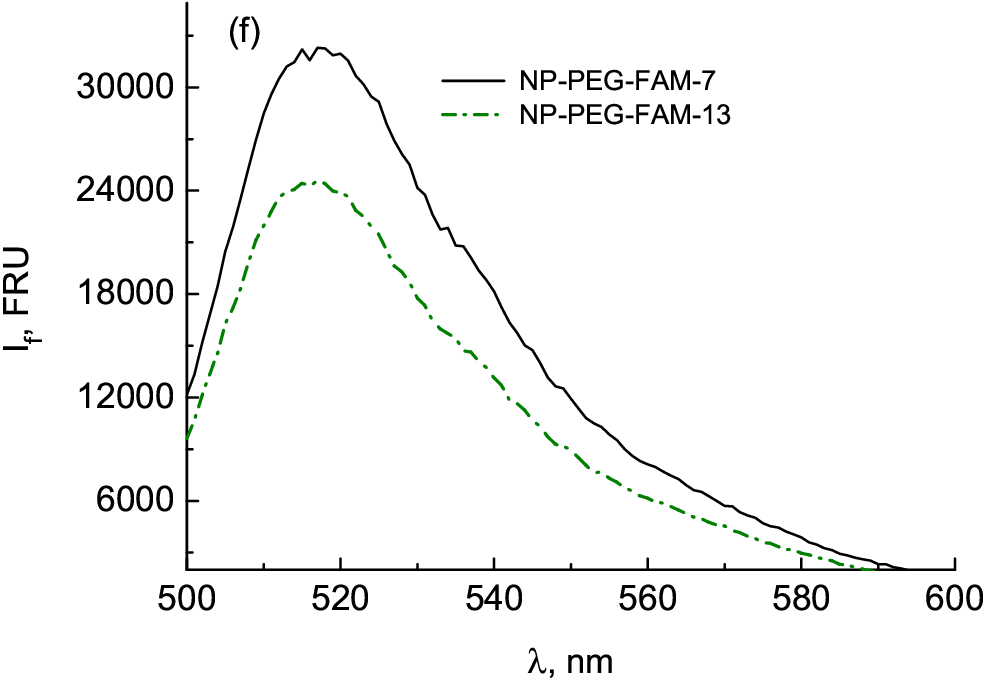}
	\caption{ Transmission electron microscopy images of (a) NP-7, (b) NP-13. Length of scale bars correspond to 100 nm. 
 (c) Electrophoretic comparison of the NP-13 (line 1), NP-PEG-FAM-13 (line 2), NP-7 (line 3), NP-PEG-FAM-7 (line 4). The NP concentration was 100 nM. (d) Optical adsorption spectra, (e) fluorescence excitation spectra, (f) fluorescence emission spectra of the obtained samples of NPs. 
 }
	\label{k3n}
\end{figure*}
 
  A suspension of NPs exhibited a characteristic surface plasmon band at 520 nm. 
    The concentration of NP-13 and NP-7 were calculated from absorbance at 520 nm using known extinction values for particles of different sizes \cite{Liu2007}.    
The ﬂuorescence intensity of samples of containing FAM-labeled NP-PEG was measured using Clariostar plate ﬂuorimeter (BMG Labtech, Ortenberg, Germany).
The nanoparticles of 7 and 13 nm shown in Fig. \ref{k3n} (a) and (b) have approximately 10$\%$ deviation in size.
It is seen in Fig. \ref{k3n} (c), that the presence of a negatively charged polyethylene glycol and FAM on the NP surface leads to an increase in the electrophoretic mobility of nanostructures. Optical adsorption, fluorescence excitation  and  fluorescence emission spectra are shown in Fig. \ref{k3n} (e), (f) and (g), respectively of obtained samples of NPs.
The maximum of the optical absorption spectra of pure NPs corresponded to 520 nm. 
Gold NPs are known to be effective quenchers of the fluorescence signal if the distance between the NPs and the fluorescent label is less than 5 nm \cite{Navarro} or if the fluorescent label is directly non-covalently attached to the surface of a gold NPs \cite{Epan}.

Hydrodynamic diameter D$_h$ and 
  $\zeta$-potential of the obtained samples of NPs are given in Table.
  
  {\bf Table}  

\begin{center}
%\begin{tabular}{ c c c }
\begin{tabular}{ |c|c|c| } 
 \hline
              & D$_h$, nm       & $\zeta$, mV \\ 
\hline
 NP-7         & 17.0$\pm$ 3.5  & -30.3$\pm$ 4.9 \\ 
 NP-PEG-FAM-7 & 36.0$\pm$ 16.0 & -48.0$\pm$ 1.0 \\  
 NP-13        & 24.1$\pm$ 18.4  & -7.6$\pm$ 1.7 \\
 NP-PEG-FAM-7 & 98.6$\pm$ 57.7  & -16.9$\pm$ 2.8 \\
 \hline
\end{tabular}
\end{center}
The hydrodynamic diameter of NP structures which determines the diffusion of nanoparticles in fluid (its dynamic characteristics) and the their $\zeta$  potential refers to the surface charge.

\subsection{For cell lines study}
A549 human lung carcinoma cells (purchased: ATCC $\#$ CCL185) were grown in Dulbecco's Modified Eagle Medium: Nutrient Mixture F12 (DMEM:F12, Sigma 
Aldrich) supplemented with 10$\%$ fetal bovine serum (GIBCO, USA), 2 mM L-glutamine, 250 mg/mL amphotericin B and 100 U/mL penicillin/streptomycin in 5 $\%$ CO$_2$. 
NCI-H23 epithelial-like lung adenocarcinoma cells (ATCC $\#$ CRL-5800) were grown in RPMI 1640 (Gibco, USA) supplemented with 10$\%$ fetal bovine serum 
(Sigma-Aldrich), 2 mM L-glutamine, 250 mg/mL amphotericin B and 100 U/mL penicillin/streptomycin in a humidified atmosphere containing 5$\%$ CO$_2$ at 37$^\circ$C. uMel1 cells were grown in 
$\alpha $-MEM medium (Sigma-Aldrich, St. Louis, MO, USA) supplemented with 20$\%$ fetal bovine 
serum (GIBCO, Thermo Fisher, Waltham, MA, USA), 

1x GlutaMAX$^{TM}$, 250 mg/mL amphotericin B 
and 100 U/mL penecillin/streptomycinc, 5
$\mu$g/mL insulin (SCI-store, Moscow, Russia), 
20 ng/mL EGF (SCI-store, Moscow, Russia), 1x 
MITO (BD Biosciences – Discovery Labware, San 
Jose, CA, USA). A549 and H23 cells were 
maintained as previously described \cite{Patrakova}. BrCCh4e-134 cells were cultivated as previously described \cite{Koval2019}. 

\subsection{ Cell’s Viability Assay}
The cells that  had  reached 30$\%$ confluence in a 96-well plate (TPP, Trasadingen, Switzerland) were treated with CAPJ or NPs and their combination. Cell viability was detected 24 hours after CAPJ irradiation using a MTT (3-[4,5-dimethylthiazol-2-yl]-2,5 diphenyl
tetrazolium bromide) test as was described previously \cite{Patrakova,Koval2017}. 

\subsection{Flow cytometry analysis of NP penetration into the cells}
Cells were growing in 96-well plates under standard conditions before the treatment with NPs/CAPJ and their combination. Before analysis cells were detached from the wells by TrypLE$^{TM}$ Express (GIBCO, Thermo Fisher Scientific, USA) treatment and washed by phosphate buffer (PBS). All analyses were performed using a FACSCantoII flow cytometer (BD Biosciences, Franklin Lakes, NJ, USA), and the data were analysed by FACSDiva Software (BD Biosciences). Cells were initially gated based on forward (FSC) versus side scatter (SSC) to exclude small debris, and ten thousand events from this population were collected. FAM-positive cell population was detected in the FITC fluorescence channel ($\lambda_{ex}$ = 491 nm, $\lambda_{em}$= 525 nm).
\subsection{Mice treatment and ethic Statement}
3H/He mice aged 6–8 weeks were obtained from the SPF vivarium of the Institute of Cytology and Genetics of the Siberian Branch of the Russian Academy of Science (SB RAS) (Novosibirsk, Russia). Mice were housed in individually ventilated cages (Animal Care Systems, Colorado, USA) in groups of 2 animals per cage with ad libitum food (ssniff, Soest, Germany) and water.
All animal experiments were carried out in compliance with the protocols and recommendations for the proper use and care of laboratory animals (EEC Directive 86/609/EEC). The study protocol was approved by the Committee on the Ethics of Animal Experiments of the Administration of SB RAS (Permit $\#$61/2 from August, 14 2020). 
A 1.0$\times$1.0~cm$^2$ area was shaved in the back region of the animals to expose the skin area. The mice were narcotized by intraperitoneal injection with a xylazine (Interchemie, The Netherlands) at a dose of 0.7 mg/kg and Zoletil 100 (Virbac, France) at a dose of 4 mg/kg, which lasts up to 20 min. The shaved areas were irradiated with CAPJ so that the distance from the animal skin to the nozzle was 25 mm. In parallel, skin temperature was measured at the place where the jet touched using a thermovision camera. After CAPJ exposure, mice are brought back to their cage in order to wake up. 

\subsection{Statistics}
Significance was determined using a two-tailed Student’s t-test. A p value of less than 0.05 was considered significant. All the error bars represent the standard deviation of the mean.

\section{CAPJ operation modes}
In our earlier study of CAPJ regimes we demonstrated that the intensity of CAPJ interaction with the target is a non-monotonic function of voltage amplitude and frequency  \cite{SCH2022,SCH2023,SCH2020L}. 
The aim of this study  is to find a CAPJ regime with optimal characteristics:  1) streamers  touch the target at each voltage cycle, 2) the current amplitude near the target I has the maximum value and 3) the temperature in the contact zone is less than 42$^\circ$C. 
We expected that the maximal cytotoxic effect of CAPJ treatment depends on the regularity of streamer propagation to the target and increases with increasing discharge current on the treated surface.

To effectively kill cancer cells, it is necessary to generate CAPJ at higher voltages since in some cases a low-energy treatment can stimulate cancer cell growth \cite{Woedtke}. At high CAPJ generation voltages, the large electric field in the streamer head activates cancer cells and promotes the penetration of nanoparticles and chemically active radicals into the cells \cite{Cheng2014,Busco,Kaneko,Shaw}. 
The concept of "activation state" was introduced by  Yan ${\it et}$ ${\it al.}$  \cite{Yan2018} and related to a significant increase in the sensitivity of CAPJ-treated cancer cells to reactive species. 
The authors hypothesized that this may be due to changes in specific pathways or expression of specific proteins in CAP-treated cells, but so far, the mechanism underlying this process is unknown.
 
The CAPJ consists of streamers that appear near the powered electrode at each voltage cycle and propagate to the target direction.  The scenario of streamers propagation outside the dielectric channel is determined by the amplitude and frequency of the voltage, which set the background plasma concentration between the nozzle and target generated by the streamers. 
Typically, some of the streamers approach the target and the rest decay near the nozzle of the device. Note that the cytotoxicity of CAPJ increases with the number of streamers reaching the treated object, since the integral rate of the radical production increases near the treated surface. 

The reason why some streamers are short and some streamers are long and approach the target is the accumulation of quasineutral plasma in  the nozzle-target gap.
The streamer can overpass this obstacle if the plasma density  in the streamer head is larger than the  quasineutral plasma density. Between voltage pulses, the plasma  decays due to recombination and drift-diffusion processes. When the plasma density decreases the next streamer can overpass the plasma cloud and reaches the target.
Depending on $\it U$ and ${\it f}_U$, the  current I over the target has the frequency N times less than the applied voltage frequency, i.e., ${\it f}_I$=${\it f}_U$/N, where N=1,2,3.... 

In Ref.\cite{SCH2020L}, we presented the map of different regimes of streamer propagation depending on the voltage amplitude and frequency, which summarized our experimental and numerical results.
The patterns of streamers  propagation (with different ${\it f}_I$=${\it f}_U$/N) are the result of self-organization of the system and depend on $\it U$ and ${\it f}_U$ (for given gas velocity, nozzle-target gap, target type).
With increasing $\it U$ and ${\it f}_U$ the  charge accumulation on the target surface forces the plasma jet to oscillate over the surface and the discharge can transit to the high current mode.
 
We found that for sinusoidal voltage CAPJ (sin CAPJ) the surface heating quickly increases with the voltage amplitude $\it U$ and frequency ${\it f}_U$.
For example, for helium gas flow rate of 6~L/min the temperature of the surface increases with the rate of 20$^\circ$C/kV for $\it U$=2.25 -- 3.8~kV and for 9~L/min with 15$^\circ$C/kV for $\it U$=2.25 -- 4.2~kV.
%(see Fig. 6 in \cite{SCH2023}).
To decrease the surface temperature, we conducted the CAPJ experiment with a controlled limitation of the  voltage shape with cutting the sinusoidal voltage in a positive half of cycles \cite{GUGIN2021}. 
For the case of clipped sinusoidal voltage,
a decrease of the current duration near the surface was seen and the surface temperature dropped by 20$\%$. By analogy with this CAPJ experiment with sinusoidal voltage, we manipulate with the pulse duration of PP voltage to avoid the overheating.

For PP CAPJ the modes with different ${\it f}_I$=${\it f}_U$/N
 were also found in the experiment and simulations. 
The background density of quasi-neutral plasma in the nozzle-target gap n$_p$ and the surface charge on the target surface $\sigma $ determine the streamer passage from the nozzle to the target. With a longer pulse duration, the streamers stay longer at the target surface, so n$_p$, and $\sigma $ are higher than with a short pulse.
\subsection{Positive-pulsed voltage CAPJ. Effect of pulse duration.}
The voltage amplitude of $\it U$=3.8 -- 4.2~kV at ${\it f}_U$=30~kHz was used for PP CAPJ generation. For the higher $\it U$ and ${\it f}_U$ the temperature  in the plasma-target contact zone was considerably higher than 42$^\circ$C.
In the experiment, the current I near the target is recorded during the CAPJ exposure to analyze the frequency of touching the target by streamers.
In numerical calculations, we track the propagation  of streamers and the current near the target over tens of voltage cycles.
Varying the voltage pulse duration $\it \tau$ from 5~$\mu$s to 16~$\mu$s allowed us to find the maximum current amplitude  at low surface heating (T$<$42$^\circ$C). 

Measurement of the current was performed on the dielectric plate and on the shaved skin of live mice with simultaneous temperature control. 
The streamer propagation characteristics were the same for these two cases, while the current amplitudes differed by 20 $\%$.
These results are useful for the translation of CAPJ treatment conditions for the therapy of tumor-bearing mice. 

Let us consider CAPJ characteristics  for the pulse duration of $\it \tau$= 7~$\mu$s and $\it \tau$=14 $\mu$s. 
 Fig.\ref{VAX} shows the applied voltage and the current measured on a shaved mouse skin during exposure to PP voltage CAPJ.
\begin{figure*}
\centering
\includegraphics[width=7.5 cm]{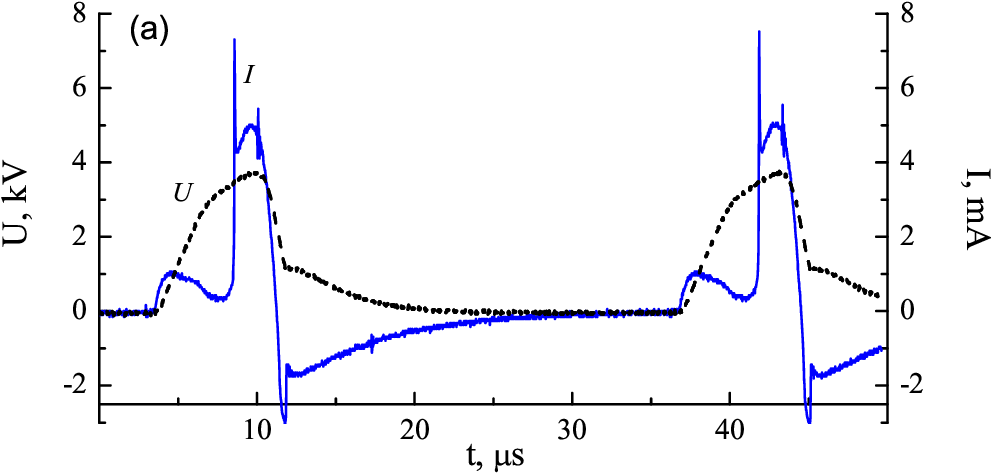}
\includegraphics[width=7.5 cm]{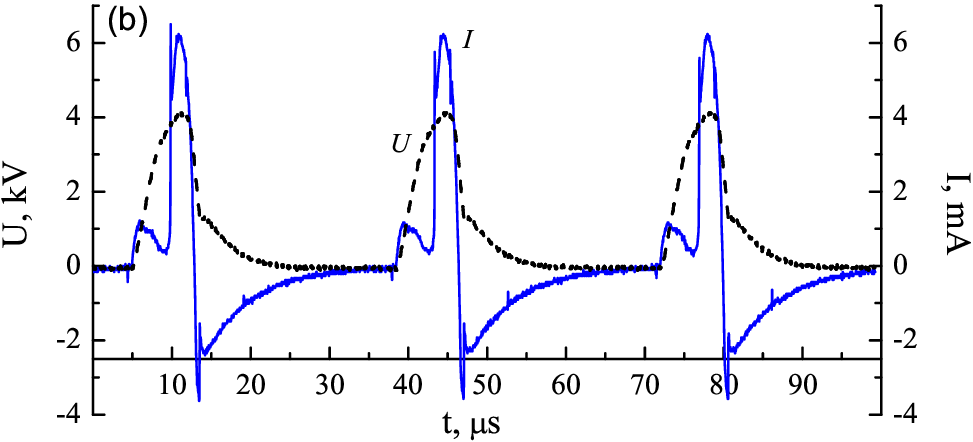}
\includegraphics[width=7.5 cm]{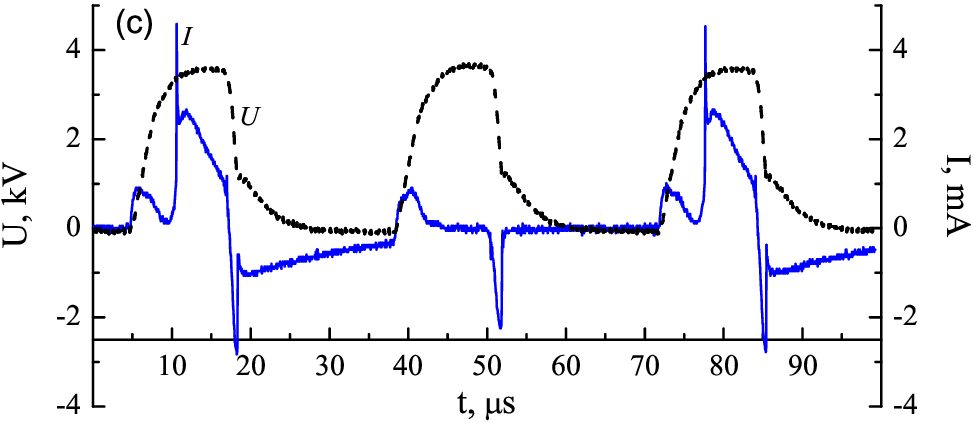}
\caption{Voltage on powered electrode  and current measured on the skin of shaved mouse exposed to CAPJ with PP voltage with ${\it f}_U$=30~kHz and (a) $\it U$=3.8~kV, $\it \tau$=7$\mu$s,  (b) $\it U$=4.2 kV, $\it \tau$=7$\mu$s and (c) $\it U$=3.8 kV, $\it \tau$=14$\mu$s. }
\label{VAX}
\end{figure*}
For $\it \tau$=7~$\mu$s, the current is registered in each voltage cycle and the current amplitude increases from 4.8~mA to 6~mA as the voltage increases from 3.8~kV to 4.2~kV (Fig.\ref{VAX}(a) and (b)). 
The maximum current for 2-3 $\mu$s was taken as the current amplitude.
For $\it \tau$=14~$\mu$s, the current frequency is half the voltage frequency 
${\it f}_I$=${\it f}_U$/2 and the current amplitude is 2.2 mA for  3.8~kV (Fig.\ref{VAX}(c)).

 In simulations, the system reaches the quasi-steady state during first three voltage cycles.  
  The calculated conduction current 
  ${\it I}_c$,  the displacement  current 
  ${\it I}_{dis}$ near the dielectric plate and z-coordinate of streamer head with time are shown in Fig.\ref{Js} for $\it \tau$=7~$\mu$s (PP-7) and 14~$\mu$s (PP-14). 
  The displacement current was calculated from the time derivative of the electric field strength near the target surface. 
     For PP-7 CAPJ the calculated currents are almost the same for each voltage cycle (Fig. \ref{Js}(a)). For PP-14 CAPJ, 
    the current near the target is two times smaller compared to PP-7 case when a streamer approached the target and is practically zero when a streamer decay inside the dielectric channel (Fig. \ref{Js}(b)).
The inserts in Fig. \ref{Js} show the z-coordinate of the streamer head (associated with the maximum ionization rate in simulations) for six voltage cycles.
As in the experiment,  
     the streamers in the PP-7 CAPJ touch the target at each voltage cycle, ${\it f}_I$=${\it f}_U$, and for PP-14 only every second streamer approaches the target, ${\it f}_I$=${\it f}_U$/2.
These calculated current amplitudes are in good agreement with measured currents in Fig.\ref{VAX}. 
On the basis of these observations, we predict the higher cytotoxicity of CAPJ with PP-7. 
  
  \begin{figure*}
\centering
\includegraphics[width=8.5 cm]{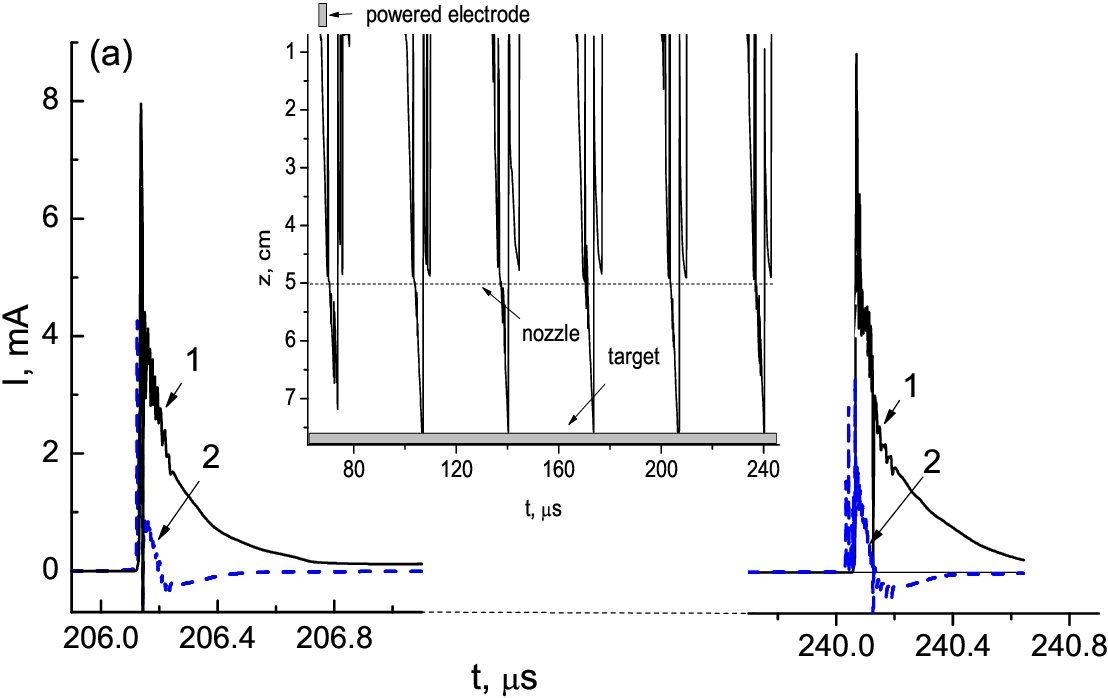}
\includegraphics[width=8.5 cm]{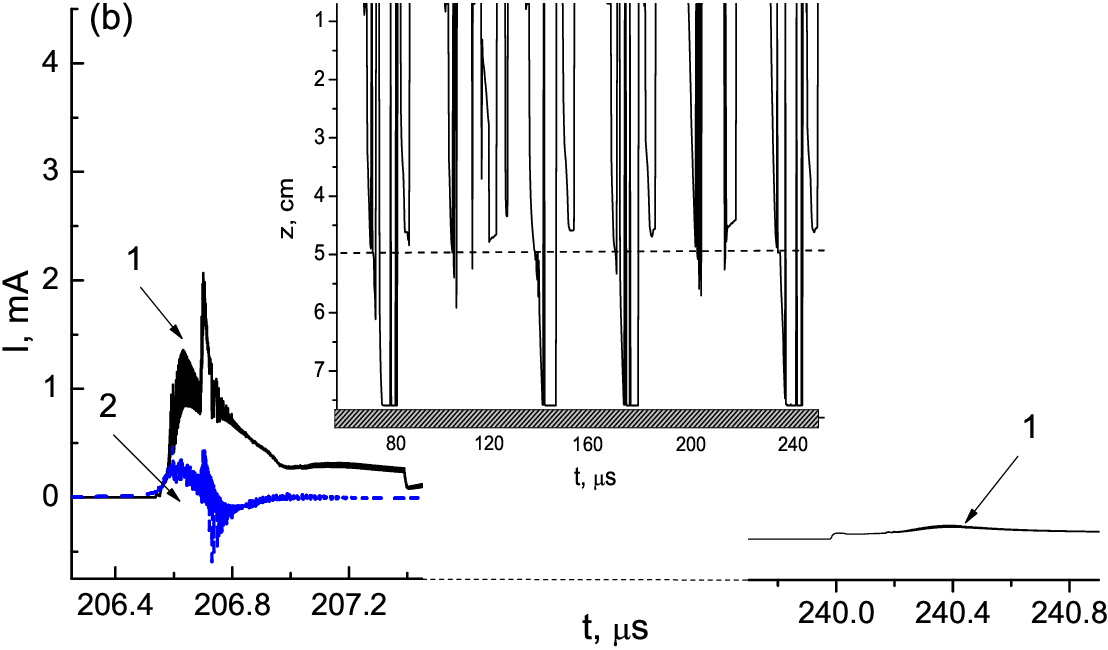}
\caption{ Calculated conduction $I_c$ (1) and displacement $I_{dis}$ (2) currents with time near the dielectric surface for PP voltage with $\it U$=4.2 kV, ${\it f}_U$=30 kHz for $\it \tau$=7$\mu$s (a) and $\it \tau$=14$\mu$s (b). Inserts show the streamer head propagation with time from the powered electrode to the target during six voltage cycles.
}
\label{Js}
\end{figure*}

\subsection{CAPJ ignited with sinusoidal voltage}
%Let us compare the currents near the surface for sin CAPJ and PP CAPJ. 
Changing the voltage amplitude we studied the different regimes of CAPJ with sinusoidal voltage  to find the mode with the maximum current  and at the temperature T$<$42$^\circ$C.
 The measurements and numerical simulations showed that the most efficient sin CAPJ mode is at $\it U$=3.3 -- 3.5~kV and frequency ${\it f}_U$=50~kHz. 
 In this mode, only every fourth streamer reached the surface, i.e. the frequency of streamer touching the target was 50/4 kHz. 
 The bio experiments confirmed that sin CAPJ  with  $\it U$=3.3 -- 3.5~kV and ${\it f}_U$=50~kHz is the most effective  for suppression of cancer cells growth.
  Fig.\ref{sin50_4} from our paper \cite{SCH2023} shows the measured current near the target, the  applied voltage, and the viability of A549 cells after exposure to sin CAPJ, at $\it U$ = 3.5 kV,  ${\it f}_U$=13 kHz and ${\it f}_U$=50 kHz.
\begin{figure*}
\centering
\includegraphics[width=6.5 cm]{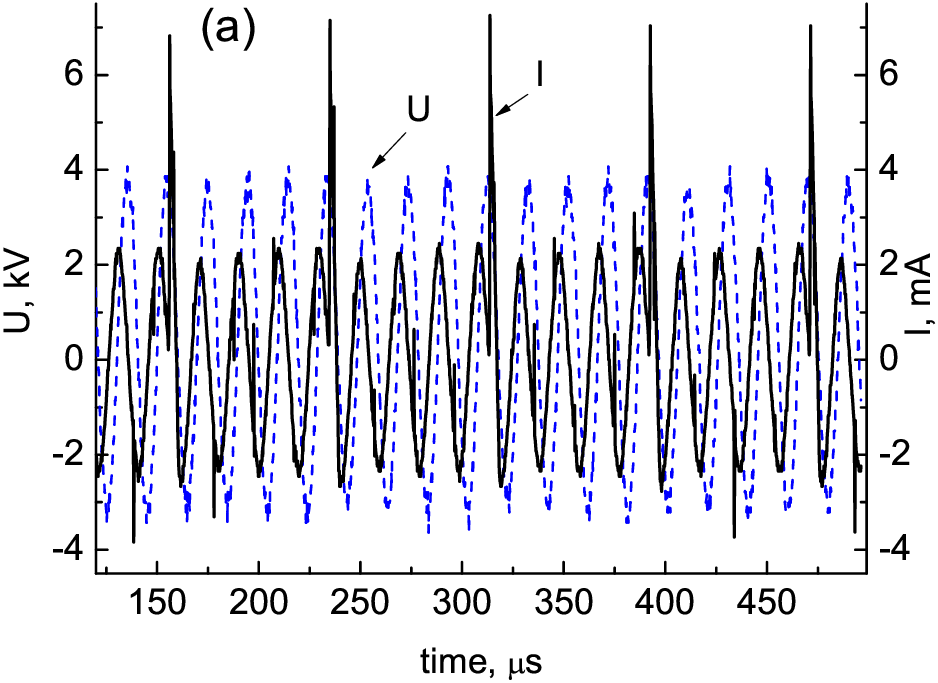}
\includegraphics[width=6.5 cm]{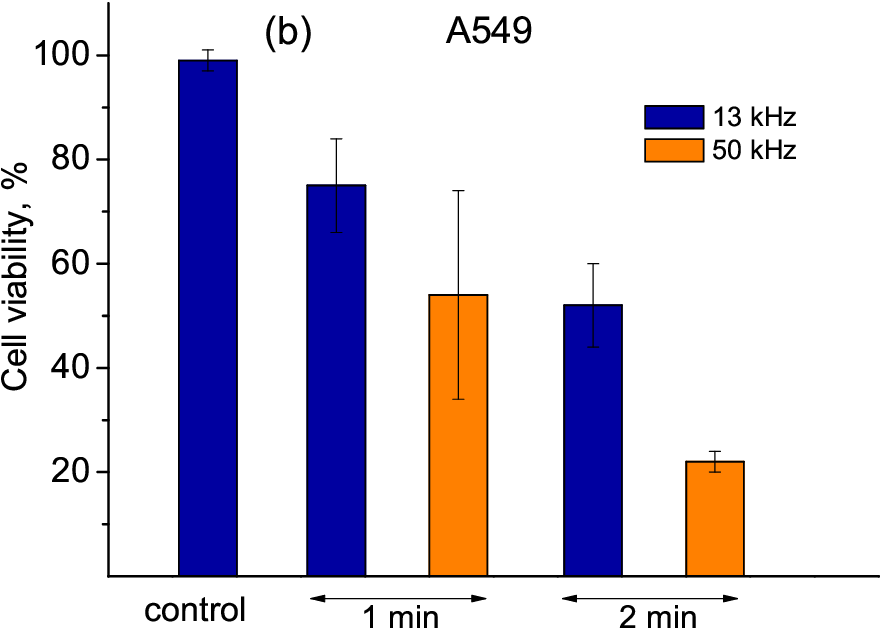}
\caption{Measured voltage (blue) and discharge current
(black) over the dielectric surface, $\it U$ = 3.5 kV, ${\it f}_U$=50 kHz  (a) and viability of A549 cells after exposure to CAPJ, $\it U$ = 3.5 kV, ${\it f}_U$=13 kHz and ${\it f}_U$=50 kHz (b) as was presented in Ref. [7]. }
\label{sin50_4}
\end{figure*}
It should be noted that this effective sin CAPJ mode is difficult to maintain in the experiment. It can suddenly transit to a regime with the higher current and strong heating due to statistical fluctuations of plasma parameters.

\section{Sensitivity of cancer cells to co-treatment with CAPJ and NP-PEG  } 

Taking into account the results of experimental and theoretical study on optimal CAPJ regimes discussed in Sec.3.1 and 3.2,  the following regimes were chosen for irradiation of the cancer cells: sin CAPJ regime  with  $\it U$=3.3~kV, ${\it f}_U$=50~kHz,  and ${\it f}_I$= of 50/4~kHz (see Fig.5 (a)), and  two PP CAPJ regimes with $\it U$=4.2 kV, ${\it f}_U$=30 kHz and  $\it \tau$=7~$\mu$s (PP-7) and  
$\it \tau$=14~$\mu$s (PP-14).  Helium flow was 9 L/min and the nozzle to target gap is 25 mm. 

 For study of cytotoxic effect of the co-treatment, we choosed  cancer cell lines belonging to different types of neoplasia: A549 lung adenocarcinoma cells, BrCCh4e-134 breast adenocarcinoma cells \cite{Koval2019} and uMel1 uveal melanoma cells.  Cell lines were chosen so that the doubling time was close in value (24$\pm$5 hours).
 The effect of co-treatment with CAPJ and gold nanoparticles covered with polyethylene glycol (NP+PEG) was analyzed  for these CAPJ regimes. Nanoparticles were added to the media with cells to a concentration of 20 nM. The diameter of pure gold nanoparticles was 13 nm, and the diameter of nanostructure NP+PEG was approximately 25$\pm$5.4~nm.
Cells were incubated with NP-PEG  for one hour and then cells were exposed to CAPJ. 
The changes in viability of treated cells were tested by MTT method 24 hours later. 
Fig.\ref{ko1} shows the viability of A549, BrCCh4e-134 and uMel1 cells.
The cytotoxic effect was considered for the following treatment cases: 1) adding NPs to media with cells without CAPJ (NP), 2) sin CAPJ  without NPs (0 NP) and 3) with NPs (20 nM NP), 4) PP-7 CAPJ without NPs (0 NP) and 5) with NPs (20 nM NP) and  6) PP-14 CAPJ without NPs (0 NP) and 7) with NPs (20 nM NP).
It is seen in Fig.\ref{ko1}, that the same trend occurs for all cell lines:  the cytotoxic effect increases in order from sin CAPJ, PP-14 CAPJ and PP-7 CAPJ.  Addition of NPs enhances the cytotoxic effect of CAPJ exposure for 1 min. 
\begin{figure*}
%\begin{figure}[ht!]
	\centering
\includegraphics[width=10. cm,keepaspectratio]{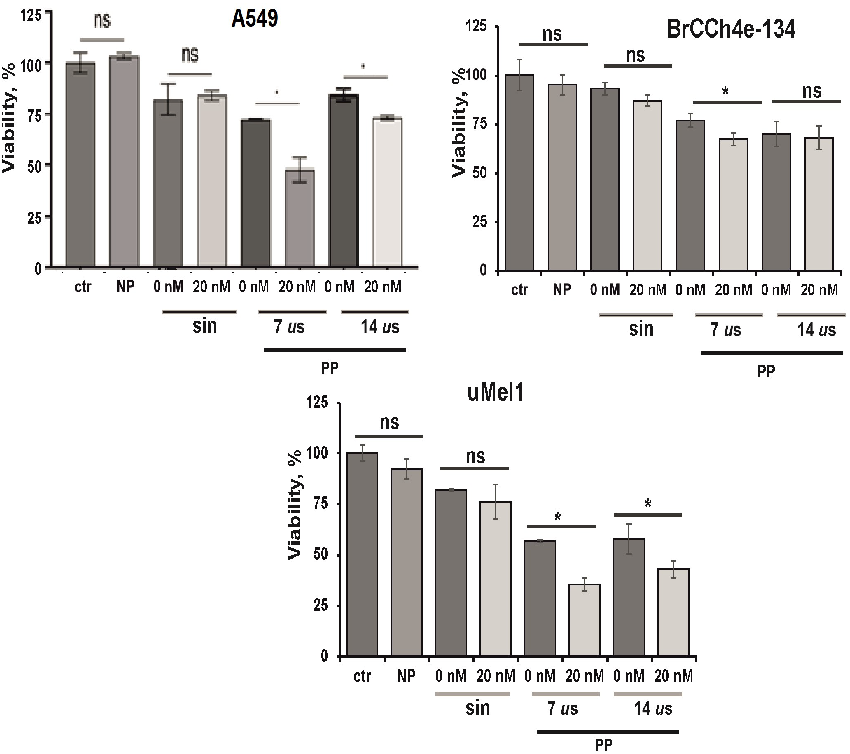}
	\caption{
 Viability of A549, BrCCh4e and uMel1 cells after co-treatment  with CAPJ and NP-PEG (20 nM). NP–PEG  were added to the cells 1 h before CAPJ treatment.  For the treatment cells were growing in the culture medium specified in Methods. The distance from the cell layer to the nozzle was 25 mm, the fluid layer over the cells was 3 mm.  MTT analysis of cell viability  was made 24 h later.  Non-treated cells were used as a positive control (ctr). 
 Data are presented as mean viability $\pm$ SD. Three replicates were performed. Statistical differences are indicated as $^\ast$ for p$<$0.05, ns - non significant (p$>$0.05).
}
\label{ko1}
\end{figure*}

%From three cell lines, BrCCh4e-134 cells were raver resistant to both CAPJ regimes with sinusoidal and positive-pulsed voltages, as well as to NP-PEG or CAPJ plus NP combination.
Among three cell lines, BrCCh4e-134 cells were relatively resistant to the
treatment with sin CAPJ and PP CAPJ, as well as to the treatment
with NP-PEG or CAPJ plus NP combination.
uMel1 cells were most sensitive to combination of PP CAPJ with NPs. This combination increased cytotoxic effect of CAPJ and the viability of cells drops to 36$\%$ for PP-7 CAPJ and to 48$\%$ for PP-14 CAPJ.
As expected from the discharge current analysis in Section 3 the PP-7 CAPJ regime is more efficient compared to the PP-14 CAPJ regime.
Analysis of viability of A549 also determined that NPs are potent to strengthen CAPJ  cytotoxic activity for both PP CAPJ regimes, but weakly affect the results for the sinusoidal regime. 

\section{Penetration of NPs into cells}

To study the effect of CAPJ exposure on NP  penetration into the cells, NP-PEG were conjugated with a fluorescent label FAM (NP-PEG-FAM) to visualize nanoparticle uptake.
A549 cells were cultured under standard conditions, and then, one hour before CAPJ treatment, NP-PEG-FAM were added to the growing cells. Treatment was performed 
with the sin CAPJ ($\it U$=2.9 kV, ${\it f}_U$=50 kHz, ${\it f}_I$=50/4~kHz, helium 9 L/min, 1 min treatment).  FAM fluorescence intensity of cells was measured by flow cytometry 90 min and 3 hours later. 
%The control was the self-fluorescence intensity of cells without NPs and CAPJ treatment. 
Fig.\ref{ko2} demonstrates the changes in 
the fluorescence intensity (P2 region) for passive uptake
of NPs without CAPJ and under CAPJ exposure. Comparison of the control sample and samples with nanoparticles without CAPJ treatment showed, that the level of passive NP entry into cells was low (2-3$\%$). Increasing the incubation time of cells with nanoparticles after CAPJ treatment from 90 min to 3 h did not increase the proportion of cells that captured NPs.
When cells were treated with CAPJ, the number of NP-positive cells increased by 10$\%$ on an average compared to passive uptake. 
Thus, direct CAPJ treatment stimulates the penetration of NP-PEG-FAM into cells. 
\begin{figure}
%\begin{figure}[ht!]
\centering
\includegraphics[width=9.9 cm,keepaspectratio]{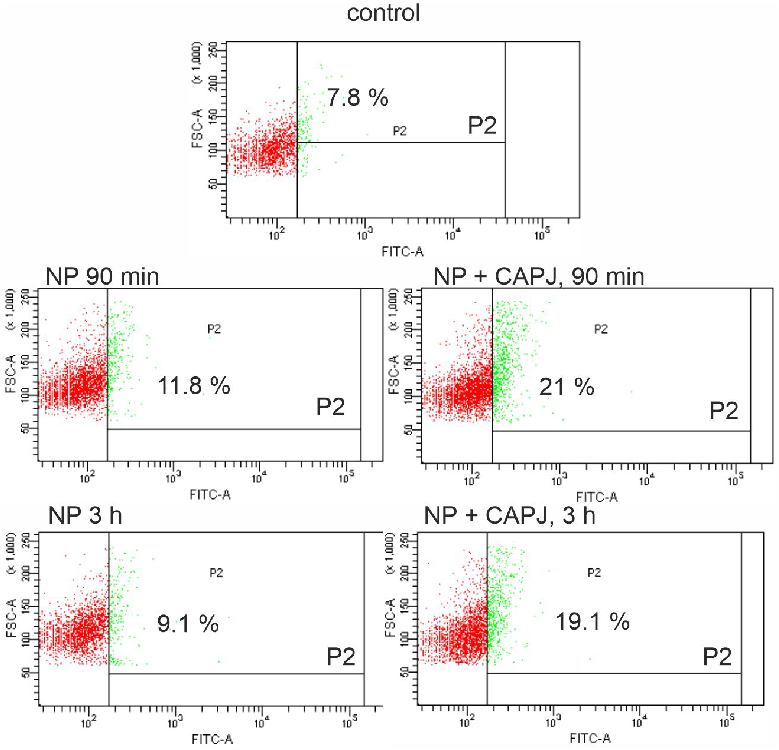}
	\caption{
Fluorescence intensity of A549 cells obtained with  flow cytometry analysis: control without NPs and CAPJ, passive uptake of NPs, uptake of NPs under CAPJ exposure. NP-PEG-FAM concentration is 20 nM,  sin CAPJ, $\it U$=2.9 kV, ${\it f}_U$=50 kHz, helium 9 L/min, 1 min treatment. NPs were added 1 h before CAPJ treatment. Flow cytometry analysis of FAM-positive cells 90 min  and 3 h post-CAPJ treatment.  P2 population was gated to determine FAM-positive cells (in FITC channel). Two replicates were performed. A representative example of analysis.
}
\label{ko2}
\end{figure}

Since increasing the time of incubation with NPs from 1.5 hours after irradiation to 3 hours did not result in an increase in NP penetration into cells, we assumed that CAPJ exposure temporarily changes the membrane permeability properties. In order to find out a time range of maximum the cell membrane permeability for NP uptake, we added NP-PEG-FAM to the cells after CAPJ treatment. A549 and H23 cells were exposed to CAPJ with sin voltage or PP voltage and NPs was added 0 - 4 hours after the treatment. Flow cytometry analysis of NP penetration into cells was performed one hour after the NP addition to cells. 
\begin{figure*}
	\centering
 \includegraphics[width=12. cm]{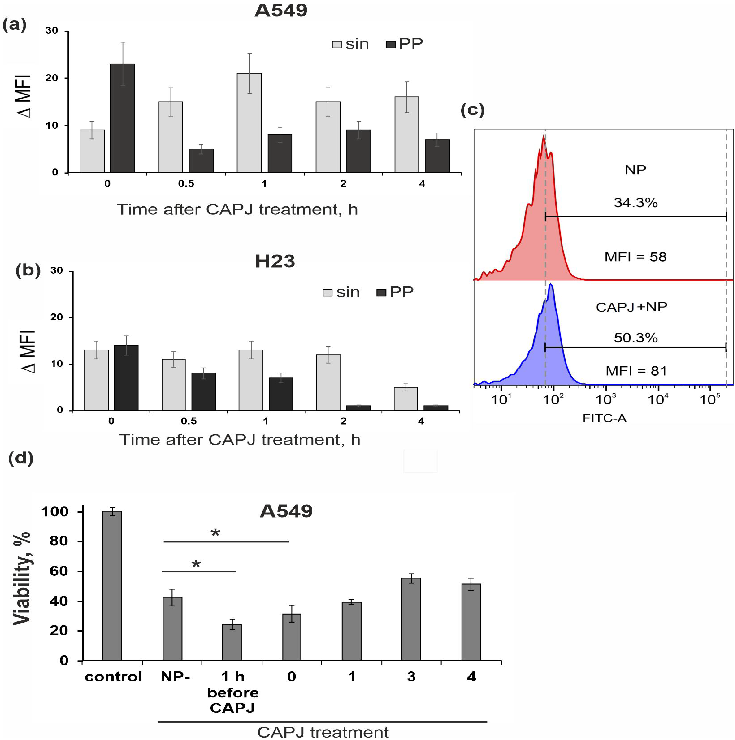}
	\caption{Dynamic of CAPJ-stimulated NP accumulation into the cells. A549 (a) and H23 (b) cells were treated with sin CAPJ  and  PP-7 CAPJ for 1 min. NP-PEG-FAM (20 nM) were added to the cells immediately (0 h) or 0.5 - 4 hours after the CAPJ treatment. Flow cytometry analysis of the cells was made one hour later. Cells treated with NPs and without CAPJ exposure were used as a negative control. Data are presented as MFI (mean fluorescence intensity) in FAM channel between CAPJ-treated NP accumulation and passive NP entering. Two replicates were performed. (c) An example of flow cytometry analysis for A549 cells treated with NP-PEG-FAM  (NP) and with NP immediately after PP CAPJ   treatment (CAPJ + NP). MFI and FITC-positive populations values ($\%$) are indicated. (d) Changes in the viability of CAPJ-exposed A549 cells under different NP addition times. A549 were treated with PP CAPJ  for 1 min. NP-PEG-FAM (20 nM) were added to the cells before, immediately (0 h) or 1 - 4 hours after the CAPJ treatment. Cells treated with NPs without CAPJ exposure (NP-) were used as a negative control. MTT analysis of the cell’s viability was made 24 h later. Non-treated cells were used as a positive control (viability 100$\%$). Three replicates were performed. Data are presented as mean viability $\pm$ SD. Statistical differences are indicated as $^\ast$ for p$<$0.05.
}
\label{ko3}
\end{figure*}
The dynamic of NP penetration was different for sin CAPJ and PP-7 CAPJ cases for both cell lines
(Fig. \ref{ko3}). 
The maximum penetration of NPs into CAPJ-treated cells was found with PP-7 voltage when NPs were added immediately after CAPJ exposure. 
When sin CAPJ voltage was applied, NPs efficiently enter the cells being added
30 min -- 2 hours after CAPJ irradiation. When treated with CAPJ, NPs penetrated more easily into A549 cells than into H23 cells.

The viability of CAPJ-treated cells is shown in Fig. \ref{ko3}(d) for different times of NP addition. The  PP-7 regime of CAPJ was applied for co-treatment with NPs.
The comparison of the penetration efficiency under different modes of NP addition (Fig. \ref{ko3}(a)) and their effect on the viability of CAPJ-treated cells (Fig. \ref{ko3}(d)) shows a direct correlation between the efficiency of nanoparticle penetration into cells and the increase of cytotoxic effect.
The maximum cytotoxic effect of combined treatment of PP-7 voltage CAPJ and NPs is achieved when NPs were added 1 hour before or immediately after CAPJ exposure.

\section{Conclusion}

The characteristics of helium cold atmospheric plasma jet generated with positive-pulsed voltage have been studied for co-treatment of cancer cells with CAPJ and gold nanoparticles. We have demonstrated the importance of optimal choice of CAPJ regime to achieve high cytotoxicity of cancer cell treatment.
In the experiment and numerical simulations, an optimal CAPJ mode was found through changing the voltage amplitude and frequency, as well as the voltage pulse duration. 
Measuring the current and temperature near the target it was shown that PP CAPJ mode with the voltage pulse duration of 7 $\mu$s (PP-7), at $\it U$=3.8 -- 4.2~kV and ${\it f}_U$=30~kHz can be the most efficient for cancer cell killing.

In bio experiments, the cytotoxic effect of   CAPJ exposure in a combination with gold nanoparticles were measured for three lines of cancer cells  of A549 lung  adenocarcinoma, BrCCh4e-134 breast adenocarcinoma,  and uMel1 uveal  melanoma.
 The treatment was performed  with CAPJ  operating in optimal  (PP-7 CAPJ) and in not-optimal (PP-14 CAPJ) modes, as well as the sin CAPJ regime.

Maximum  cytotoxic effect was observed
for  CAPJ mode excited with PP-7 CAPJ case  that confirmed the results of discharge current analysis in Sec.3. The effect of treatment increased, with lowest observed for sin CAPJ case, followed by PP-14 CAPJ and PP-7 CAPJ cases.  
Addition of NPs+PEG to the medium with cells followed by CAPJ exposure showed that NP uptake  by cells is enhanced by CAPJ by 13$\%$ compared to passive NP uptake  of  2-3$\%$. The addition of NPs enhanced the cytotoxic effect of CAPJ exposure, the magnitude of which correlated with the mode of maximal NP penetration.
  BrCCh4e-134 cells were found to be more resistant to co-treatment,  whereas uMel1 and A549 cells were more sensitive to CAPJ and NPs+PEG exposure. 

  On the basis on experimental and theoretical study of sinusoidal and positive-pulsed voltage CAPJ regimes, we  have selected modes of CAPJ exposure of cancer cells when gold NPs effectively penetrate into cells and reduce their viability. To enhance the cytotoxic activity and increase the targeting effect, the nanoparticles can carry in the therapeutic agents and target molecules on the next stages of the study. Since the  proposed modes of CAPJ operation has been characterized in the experiment and numerical simulations, and the thermal safety has been demonstrated it allows us to translate these regimes for the treatment of the animals with tumors.

 \ack 
The authors gratefully acknowledge financial support from Russian Science
Foundation grant number 22-49-08003.
The research on uMel1 cells cultivation was funded by Russian Science Foundation grant number 23-14-00285.
{}


\section*{References}


\begin{thebibliography}{42}
\bibitem{Zivani2023}
$\breve{Z}$ivani M., Espona-Noguera A., Lin A., Canal C. // Adv. Sci. 2023 , V.10(8), p.2205803. doi:10.1002/advs.202205803
\bibitem{Limanowski}
Limanowski, R.; Yan, D.; Li, L.; Keidar, M. Preclinical Cold Atmospheric Plasma Cancer
Treatment. Cancers 2022, 14, 3461. https://doi.org/10.3390/cancers14143461
\bibitem{Lya2023}
 Lya L., ChengaX., Murthya S., Zhuang T., Jones O., Basadonna G., Keidar M., Canady J. //  Clinical Plasma Medicine, 2023, in press.
\bibitem{Woedtke}
von Woedtke, T.: Schmidt, A.; Bekeschus, S.; Wende, K.; Weltmann, K.D, Plasma medicine: a field of applied redox biology.  {\it In Vivo} {\bf 2019}, {\it 33}, 1011-1026.
DOI:10.21873/invivo.11570 

\bibitem{Yan2018}
Yan, D.; Xu, W.; Yao, X.; Lin L.; Sherman J.; Keidar M. 
The Cell Activation Phenomena in the Cold Atmospheric Plasma Cancer Treatment.
{\it Sci. Rep.} {\bf 2018} {\bf 8}, 15418. doi:10.1038/s41598-018-33914-w.

\bibitem{SCH2022}
Schweigert, I.V.; Zakrevsky, D.E.; Gugin, P.P.; Milakhina, E.V.; 
Biryukov, M.M.; Keidar, M.; Koval, O.A. 
Effect of voltage pulse duration on electrophysical and thermal
characteristics of cold atmospheric plasma jet.
{\it Plasma Sources Sci. Technol.} {\bf 2022}, {\it 31}, 114004. 
DOI:10.1088/1361-6595/aca120

\bibitem{SCH2023}
Schweigert, I.V.; Zakrevsky, D.E.; Milakhina, E.V.; Gugin, P.P.; 
Biryukov, M.M.; Patrakova, E.A.; Troitskaya, O.S.; Koval, O.A. 
Characteristics of cold atmospheric plasma jet when excited by 
sinusoidal and positive pulse voltages for medical applications.
{\it Plasma Physics Reports} {\bf 2023}, {\it 49}(5), 595-601.

\bibitem{Kim2009}
Kim, G.G.; Kim, G.J.; Park, S.R.; Jeon, S.M.; Seo, H.J.; Iza, F.; Lee, J.K.
Air plasma coupled with antibody-conjugated nanoparticles: a new weapon
against cancer.
{\it J. of Phys. D: Applied Physics} {\bf 2009}, {\it 42}, 032005.
8\bibitem{Kim2011}
Kim G.J. et al. Targeted Cancer Treatment Using Anti-EGFR and-TFR Antibody-Conjugated Gold Nanoparticles Stimulated by Nonthermal Air Plasma // Plasma Medicine. 2011. Vol. 1, № 1. 45–54 p.

\bibitem{Cheng2014}
Cheng, X.Q.; Murphy, W.; Recek, N.; Yan, D.; Cvelbar, U.; 
Vesel, A.; Mozetic, M.; Canady, J.; Keidar, M; Sherman J.H.
Synergistic effect
of gold nanoparticles and cold plasma on glioblastoma cancer therapy. 
{\it J. of Phys. D: Applied Physics} {2014}, {\it 47}, 335402.
DOI:10.1088/0022-3727/47/33/335402

\bibitem{Choi2015}

Choi, B.B.; Kim, M.S.; Kim, U.K.; Hong, J.W.; Lee,  H.J.; Kim, G.C.
Targeting NEU protein in melanoma cells with non-thermal atmospheric pressure 
plasma and gold nanoparticles.
{\it J Biomed Nanotechnol.} {\bf 2015}, {\it 11}, 900-905.

DOI:10.1166/jbn.2015.1999

\bibitem{Irani2015}
Irani S. et al. Induction of growth arrest in colorectal cancer cells by cold plasma and gold nanoparticles // Archives of Medical Science. 2015. Vol. 6. P. 1286–1295.
\bibitem{Choi2017}

Choi B.B.R. et al. Selective Killing of Melanoma Cells With Non-Thermal Atmospheric Pressure Plasma and p-FAK Antibody Conjugated Gold Nanoparticles // Int J Med Sci. 2017. Vol. 14, № 11. P. 1101–1109.
\bibitem{Kim2017}

Kim W. et al. Selective uptake of epidermal growth factor-conjugated gold nanoparticle (EGF-GNP) facilitates non-thermal plasma (NTP)-mediated cell death // Sci Rep. 2017. Vol. 7, № 1. P. 10971.
\bibitem{Jawaid2020}
Jawaid P. et al. Small size gold nanoparticles enhance apoptosis-induced by cold atmospheric plasma via depletion of intracellular GSH and modification of oxidative stress // Cell Death Discov. 2020. Vol. 6, № 1. P. 83.

\bibitem{Kaushik2016}
Kaushik N.K. et al. Low doses of PEG-coated gold nanoparticles sensitize solid tumors to cold plasma by blocking the PI3K/AKT-driven signaling axis to suppress cellular transformation by inhibiting growth and EMT // Biomaterials. 2016. Vol. 87. P. 118–130.

\bibitem{Al-Harbi2020}
 Al-Harbi N.S. et al. Effect of naked and PEG-coated gold nanoparticles on histopathology and cytokines expression in rat liver and kidneys // Nanomedicine. 2020. Vol. 15, № 3. P. 289–302.
\bibitem{Schmidt}
Schmidt, A.; Liebelt, G.; Striesow, G.; Freunda, E.; von Woedtkea, T.; 
Wende, K.; Bekeschus S.
The molecular and physiological consequences of cold plasma treatment in
murine skin and its barrier function. 
{\it Free Radical Biology and Medicine} {\bf 2020}, {\it 161}, 32-49.

DOI:10.1016/j.freeradbiomed.2020.09.026 


\bibitem{Lademann}
Lademann, J.; Patzelt, A.; Richter, H.; Lademann, O.; Baier G.; 
Breucker L.; Landfester K.
Nanocapsules for drug delivery through the skin barrier by tissue-tolerable
plasma.
{\it Laser Physics Letters} {\bf 2013}, {\it 10}, 083001.

DOI:10.1088/1612-2011/10/8/083001


\bibitem{OLademann}
Lademann, O.; Richter, H.;, Kramer, A.; 
Patzelt, A.; Meinke, M.C.; Graf, C.; Gao, Q.; Korotianskiy, E.; Ruhl, E.;
Weltmann. K.-D.
Stimulation of the penetration of particles into the skin by plasma tissue
interaction.
{\it Laser Physics Letters} {\bf 2011}, {\it 8}, 758-764.

DOI:10.1002/lapl.201110055


\bibitem{OLademann1}
Lademann, O.; Richter H.; Meinke, M.C.; 
Patzelt, A.; Kramer, A.; Hinz, P.; Weltmann, K.-D.; Hartmann, B.; Koch S.
Drug delivery through the skin barrier 
enhanced by treatment with tissuetolerable plasma. 
{\it Experimental Dermatology} {\bf  2011}, {\it 20}, 488-490.

DOI:10.1111/j.1600-0625.2010.01245.x

\bibitem{Busco}
Busco, G.; Robert, E.; Chettouh-Hammas, N.; Pouvesle, J.-M.; Grillon, C.
The emerging potential of cold atmospheric plasma in skin biology.
{\it Free Radical Biology and Medicine} {\bf 2020}, {\it 161}, 290-304.

DOI:10.1016/j.freeradbiomed.2020.10.004 


\bibitem{Kaneko}
Kaneko, T.; Sasaki, S.; Hokari, Y.; Horiuchi, S.;  Honda, R.; Kanzaki, M.
Improvement of cell membrane permeability
using a cell-solution electrode for generating atmospheric-pressure plasma.
{\it Biointerphases} {2015}, {\it 10}, 029521.

DOI:10.1116/1.4921278



\bibitem{Shaw}
Shaw, P.; Kumar, N.; Hammerschmid, D.;  Privat-Maldonado, A.;
Dewilde, S.; Bogaerts, A.
Synergistic effects of melittin and
plasma treatment: a promising approach for cancer therapy.
{\it Cancers} {\bf 2019}, {\it 11}, 1109

DOI:10.3390/cancers11081109 

\bibitem{He2018}
He, Z.;  Liu, K.;  Manaloto, E.; Casey, A.; Cribaro, G.P.; Byrne, H.J.; 
Tian, F.; Barcia, C.; Conway, G.E.; Cullen, P.J.; Curtin. J.F.
Cold atmospheric plasma induces ATP-dependent endocytosis of nanoparticles 
and synergistic U373MG cancer cell death.
{\it Sci Rep} {\bf 2018}, {\it 8}, 5298.

DOI:10.1038/s41598-018-23262-0


\bibitem{Cheng2015}
Cheng,  X. Q.; Rajjoub, K.;  Sherman, J.; Canady, J.; Recek, N.; 
Yan, D.; Bian, K.; Murad, F.; Keidar, M.
Cold plasma accelerates the uptake of gold nanoparticles into glioblastoma
cells.
{\it Plasma Processes and Polymers} {\bf 2015}, {\it 12}, 1364-1369.

DOI:10.1002/ppap.201500093

\bibitem{He2020}
He, Z.; Liu, K.; Scally, L.; Manaloto, E.; Gunes, S.; Ng, S.W.; Maher, M.; 
Tiwari, B.; Byrne, H.J.; Bourke, P.; Tian, F.; Cullen, P.J.; Curtin, J.F.
Cold atmospheric plasma stimulates clathrin-dependent endocytosis to repair 
oxidised membrane and enhance uptake of nanomaterial in glioblastoma
multiforme cells.
{\it Sci. Rep.} {\bf 2020}, {\it 10}, 6985.

DOI:10.1038/s41598-020-63732-y

\bibitem{Jawaid}
Jawaid, P.; Rehman, M.U.; Zhao, Q.L.; Misawa, M.; Ishikawa, K.; 
Hori, M.; Shimizu, T.; Saitoh, J.; Noguchi, K.; Kondo, T.
Small size gold nanoparticles enhance apoptosis-induced by cold atmospheric
plasma via depletion of intracellular GSH and modification of oxidative stress. 
{\it Cell Death Discov.} {\bf 2020}, {\it 6}, 83.

DOI:10.1038/s41420-020-00314-x





\bibitem{Bekeschus2021}
Bekeschus, S. 
Combined toxicity of gas plasma treatment
and nanoparticles exposure in melanoma cells in vitro.
{\it Nanomaterials} {\bf 2021}, {\it 11}, 806.
DOI:10.3390/nano11030806
\bibitem{SCH2019}
Schweigert, I.V.; Zakrevsky, D.E.; Gugin, P.P.; Yelak, E.V.; 
Golubitskaya, E.A.; Troitskaya, O.S.; Koval, O.A.
Interaction of cold atmospheric argon and helium plasma jets with bio-target with grounded substrate beneath.
{\it Appl. Sci.} {\bf 2019}, {\it 9}(21), 4528.
DOI:10.3390/app9214528
%\bibitem{GUGIN2019}
%P.P. Gugin et al. Investigation of hydroxyl group radicals' generation at interaction of a cold atmospheric plasma jet with an environment, 15th International Conference "Gas Discharge Plasmas and Their Applications" GDP 2021, September 5-10, 2021. Abstracts p.191].

\bibitem{SCH2020L}
Schweigert, I.V.; Alexandrov, A.L.; Zakrevsky, D.E.
Self-organization of touching-target current with ac voltage in atmospheric 
pressure plasma jet for medical application parameters.
{\it Plasma Sources Sci. Technol.} {\bf 2020}, {\it 29}, 12LT02.

DOI:10.1088/1361-6595/abc93f

\bibitem{SCH2022F}
Schweigert, I.V.; Zakrevsky, D.E.;  Milakhina, E.V.; Gugin, P.P.;
 Biryukov M.M.; Patrakova, E.A.; Koval, O.A.
A grounded electrode beneath dielectric targets, including cancer cells, enhances the impact of cold atmospheric plasma jet. {\it Phys. Control. Fusion} {\bf 2022}, {\it 64}, 044015.
 https://doi.org/10.1088/1361-6587/ac53f1
 

%\bibitem{SCH2020L}
%Schweigert, I.V.; Alexandrov, A.L.; Zakrevsky, D.E.
%Self-organization of touching-target current with ac voltage in atmospheric pressure plasma jet for medical application parameters. {\it Plasma Sources Sci. Technol.} {\bf 2020}, {\it 29}, 12LT02. DOI:10.1088/1361-6595/abc93f

\bibitem{SCH2019_1}
Schweigert I.; Vagapov S.; Lin L.; Keidar M. 
Enhancement of atmospheric plasma jet-target interaction
with an external ring electrode.  {\it J. Phys. D: Appl. Phys.} {\bf 2019}
{\it 52}, 295201

\bibitem{Jana2001}

Jana N. R.; Gearheart L.; Murphy C. J. 
Seeding growth for size control of 5-40 nm diameter gold nanoparticles.
{\it Langmuir} {\bf 2001}, {\it 17(22)}, 6782–6786.
DOI:10.1021/la0104323

\bibitem{Murphy2004}
Murphy, D.; Eritja, R.; Redmond, G. 
Monitoring denaturation behaviour and comparative stability of DNA triple 
helices using oligonucleotide gold nanoparticle conjugates.
{\it Nucleic Acids Res.} {\bf 2004}, {\it 32}(7), e65. 

DOI:10.1093/nar/gnh065
\bibitem{Bartczak}
D. Bartczak and A.G. Kanaras Preparation of Peptide-Functionalized Gold Nanoparticles Using One Pot EDC/Sulfo-NHS Coupling. Langmuir 2011, 27, 10119–10123, doi:10.1021/la2022177.
\bibitem{Liu2007}
Liu, X.; Atwater, M.; Wang, J.; Huo, Q. 
Extinction coefficient of gold nanoparticles with different sizes and 
different capping ligands.
{\it Colloids Surf B Biointerfaces} {\bf 2007}, {\it 58}, 3-7.

DOI:10.1016/j.colsurfb.2006.08.005

\bibitem{Navarro}
Navarro, J.R.G.; Lerouge, F. 
From gold nanoparticles to luminescent nano-objects: experimental aspects 
for better gold-chromophore interactions. 
{\it Nanophotonics} {\bf 2017}, {\it 6}, 71-92.

DOI:10.1515/nanoph-2015-0143


\bibitem{Epan}
Epanchintseva, A.V.; Gorbunova, E.A.; Ryabchikova, E.I.; Pyshnay, I.A.; 
Pyshnyi, D.V.
Effect of fluorescent labels on DNA affinity for gold nanoparticles. 
{\it Nanomaterials} {\bf 2021}, {\it 11}, 1178.

DOI:10.3390/nano11051178
\bibitem{Patrakova}
Patrakova, E.; Biryukov, M.; Troitskaya, O.; Gugin, P.; Milakhina, E.; 
Semenov, D.; Poletaeva, J.; Ryabchikova, E.; Novak, D.; Kryachkova, N.; 
Polyakova, A.; Zhilnikova, M.; Zakrevsky, D.; Schweigert, I.; Koval O.
Chloroquine enhances death in lung adenocarcinoma A549 cells exposed to cold
atmospheric plasma jet. 
{\it Cells} {\bf 2023}, {\it 12}, 290. 

DOI:10.3390/cells12020290

\bibitem{GUGIN2021}
Gugin, P.; Zakrevsky, D.; Milakhina, E.
Electrophysical and thermal parameters of atmospheric pressure plasma jet in helium under excitation by sinusoidal and pulse voltage.
{\it Tech. Phys. Lett.} {\bf 2021},{\it 47} 41–44, in Russian. 
\bibitem{Koval2019}
Koval, O.A.; Subrakova, V.G.; Nushtaeva, A.A.; Belovezhets, T.N.; Troitskaya, O.A.; 
Ermakov, M.S.; Varlamov, M.E.; Chikaev, A.N.; Kuligina, E.V.; Kulemzin, S.V.; 
Gorchakov, A.A.; Taranin, A.V.;, Richter, V.A. 
CAR-dependent anti-metastatic activity of modified NK cell line YT. 
{\it Genes and Cells} {\bf 2019}, {\it 14}(4), 66-71. 

DOI:10.23868/201912034 

%\bibitem{Frens}
%Frens, G. Controlled Nucleation for the Regulation of the Particle Size in Monodisperse Gold Suspensions. Nat. Phys. Sci. 1973, 241, 20–22, doi:10.1038/physci241020a0


%\bibitem{Bartczak}
%D. Bartczak and A.G. Kanaras Preparation of Peptide-Functionalized Gold Nanoparticles Using One Pot EDC/Sulfo-NHS Coupling. Langmuir 2011, 27, 10119–10123, doi:10.1021/la2022177.

%\bibitem{Patrakova}
%Patrakova, E.; Biryukov, M.; Troitskaya, O.; Gugin, P.; Milakhina, E.; Semenov, D.; Poletaeva, J.; Ryabchikova, E.; Novak, D.; Kryachkova, N.; Polyakova, A.; Zhilnikova, M.; Zakrevsky, D.; Schweigert, I.; Koval O.Chloroquine enhances death in lung adenocarcinoma A549 cells exposed to cold atmospheric plasma jet. {\it Cells} {\bf 2023}, {\it 12}, 290. DOI:10.3390/cells12020290


\bibitem{Koval2017}
Koval, O.; Kochneva, G.; Tkachenko, A.; Troitskaya, O.; Sivolobova, G.; 
Grazhdantseva, A.; Nushtaeva, A.; Kuligina, E.; Richter, V.
Recombinant vaccinia viruses coding transgenes of apoptosis-inducing proteins 
enhance apoptosis but not immunogenicity of 
infected tumor cells.
{\it Biomed Res Int.} {\bf 2017}, {\it 2017}, 3620510.

DOI:10.1155/2017/3620510 

 


\end{thebibliography}
\end{document}